\documentclass[aps,showpacs,preprintnumbers,amsmath, amssymb]{revtex4}

\usepackage{float}
\usepackage{graphics,epsfig}
\usepackage{graphicx}
\usepackage{dcolumn}
\usepackage{bm}

\begin{document}
\baselineskip=0.8 cm

\title{{\bf Holographic superconductors in the Born-Infeld  electrodynamics }}

\author{Jiliang Jing}
\email{jljing@hunnu.edu.cn}
 \affiliation{ Institute of Physics and
Department of Physics, Hunan Normal University,  Changsha, Hunan
410081, P. R. China \\ Key Laboratory of Low Dimensional Quantum
Structures \\ and Quantum Control of Ministry of Education, Hunan
Normal University, Changsha, Hunan 410081, P. R. China}

\author{Songbai Chen}
\email{csb3752@163.com} \affiliation{ Institute of Physics and
Department of Physics, Hunan Normal University,  Changsha, Hunan
410081, P. R. China \\ Key Laboratory of Low Dimensional Quantum
Structures \\ and Quantum Control of Ministry of Education, Hunan
Normal University, Changsha, Hunan 410081, P. R. China}

\vspace*{0.2cm}
\begin{abstract}
\baselineskip=0.6 cm
\begin{center}
{\bf Abstract}
\end{center}

We study the effects of the Born-Infeld electrodynamics on the
holographic superconductors in the background of a Schwarzschild AdS
black hole spacetime. We find that the presence of Born-Infeld scale
parameter decreases the critical temperature and the ratio of the
gap frequency in conductivity to the critical temperature for the
condensates. Our results means that it is harder for the scalar
condensation to form in the Born-Infeld electrodynamics.

\end{abstract}

 \maketitle
\newpage
\section{Introduction}

In the last century, Maldacena \cite{ads1} has proposed an exact
correspondence between gravity theory in a $(d+1)$-dimensional anti
de Sitter (AdS) spacetime and a conformal field theory living on the
$d$-dimensional boundary. This is so-called AdS/CFT correspondence,
which now is a powerful tool to understand the strong coupled gauge
theories \cite{ads2,ads3,ads4}. In recent years, this holographic
correspondence has been applied extensively in the condensed matter
physics including superconductivity \cite{Hs0,Hs01,Hs02,Hs03} and
superfluid \cite{Hsf0,Hsf01}. The first model for holographic
superconductors in the AdS black hole spacetime is proposed in
\cite{Hs0}. The model consists of a system with a black hole and a
charged scalar field, in which the black hole admits scalar hair at
temperature $T$ smaller than a critical temperature $T_c$. According
to the AdS/CFT correspondence, the emergence of a hairy AdS black
hole means the formation of a charged scalar condensation in the
dual CFTs \cite{Hsa1}. This indicates that the expectation value of
charged operators undergoes the U(1) symmetry breaking and then the
second order phase transition is occurred. The electrical conductive
for the condensations has been calculated through the fluctuations
of the vector potential, and it is shown that the condensation has
zero electrical direct current resistance, which is the same as that
obtained in the Bardeen-Cooper-Schrieffer (BCS) theory \cite{bcs}.
This has triggered many people to study holographic superconductors
in the various theories of gravity
\cite{a0,a01,a1,a2,a3,a4,a401,a40,a402,a41,a42,
a5,a6,a7,a8,a8d1,GT,a81,a9,a10,a100,
a11,a12,a13,a14,a15,a151,a16s,a17}. The holographic superconductors
in Einstein-Gauss-Bonnet gravity have been studied in \cite{a0,a401}
and it is found that the higher curvature corrections make
condensation harder. The models for the holographic superconductors
in the Ho\v{r}ava-Lifshitz gravity have been investigated in
\cite{a40,a402,a41,a42}. The properties of holographic
superconductors have been considered in the string/M
theory\cite{a5,a6,a7,a8,a8d1} and in the zero temperature limit
\cite{a81,a9,a10,a100}, respectively. These results can help us to
understand more about the holographic superconductor in the
asymptotical AdS black holes.

All of the above models mentioned for the holographic
superconductors are in the frame of Maxwell electromagnetic theory.
It is of interest to investigate holographic superconductor in the
nonlinear electromagnetic generalization. One of the important
nonlinear electromagnetic theories is Born-Infeld electrodynamics
\cite{BI}, which was proposed in 1934 to avoid the infinite self
energies for charged point particles arising in Maxwell theory. The
Born-Infeld electrodynamics displays good physical properties
including the absence of shock waves and birefringence. Among all
electromagnetic theories, there are only two theories that have no
birefringence: Born-Infeld electrodynamics and Maxwell
electrodynamics. Moreover, Born-Infeld theory is singled out among
all nonlinear electromagnetic theories for having electric-magnetic
duality invariance \cite{BI0}. In the string theory, Born-Infeld
action can also describe gauge fields on a D-brane which arises from
attached open strings \cite{BI1}. Combining Born-Infeld
electrodynamics with general relativity, a spherically symmetric
black hole solution was obtained in \cite{BIbh}. The physical
properties of the black hole have been studied extensively in recent
years \cite{Bbh1,Bbh2,Bbh3}. In this paper, we will investigate the
holographic superconductor in planner AdS black hole in the frame of
Born-Infeld electrodynamics. The main purpose in this paper is to
see what effect of Born-Infeld scale parameter on the holographic
superconductor in this asymptotic AdS black hole.

This paper is organized as follows. In Sec. II, we give the basics
equations and study numerically holographic superconductors in the
Born-Infeld electrodynamics. Our results show that the scalar
condensation is harder to form than in the Maxwell electrodynamics.
In Sec. III, we ignore the backreaction of the dynamical matter
field on the spacetime metric and calculate the electrical
conductivity of the charged condensates. Finally in the last section
we will include our conclusions.

\section{The condensate of charged operators}
The metric of a planner Schwarzschild AdS black hole is
\begin{eqnarray}
ds^2=-f(r)dt^2+\frac{1}{f(r)}dr^2+r^2(dx^2+dy^2),\label{m1}
\end{eqnarray}
with
\begin{eqnarray}
f(r)=\frac{r^2}{L^2}-\frac{2M}{r},
\end{eqnarray}
where $L$ is the radius of AdS and $M$ is the mass of black hole.
The Hawking temperature is
\begin{eqnarray}
T_H=\frac{3r_H}{4\pi L^2},
\end{eqnarray}
where $r_H$ is the event horizon of the black hole.

Let us now consider an electric field and a charged complex scalar
field in the background of a Schwarzschild-AdS black hole. The
Lagrangian can be expressed as
\begin{eqnarray}
\mathcal{L}=\mathcal{L}_{BI}-
|\nabla_{\mu}\psi-iqA_{\mu}\psi|^2-m^2\psi^2,
\end{eqnarray}
where $\psi$ is a charged complex scalar field, $\mathcal{L}_{BI}$
is Lagrangian density of the Born-Infeld electrodynamics
\begin{eqnarray}
\mathcal{L}_{BI}=\frac{1}{b}\bigg(1-\sqrt{1+\frac{b F}{2}}\bigg).
\end{eqnarray}
Here $F\equiv F_{\mu\nu}F^{\mu\nu}$ and $F_{\mu\nu}$ is the
nonlinear electromagnetic tensor which satisfies Born-Infeld
equation
\begin{eqnarray}
\partial_{\mu}\bigg(\frac{\sqrt{-g}F^{\mu\nu}}{\sqrt{1+\frac{b F}{2}}}\bigg)=J^{\nu}.
\end{eqnarray}
The scale parameter $b$ indicates the difference between Born-Infeld
and Maxwell electrodynamics. As $b$ tends to zero, the Lagrangian
$\mathcal{L}_{BI}$ approaches to $-\frac{1}{4}F_{\mu\nu}F^{\mu\nu}$,
and the Einstein-Maxwell theory is recovered. Assuming that these
fields are weakly coupled to gravity, we can neglect their
backreactions on the metric. As in \cite{Hs0}, we adopt to the
ansatz
\begin{eqnarray}
A_{\mu}=(\phi(r),0,0,0),\;\;\;\;\psi=\psi(r),
\end{eqnarray}
and then find that the equations of motion for the complex scalar
field $\psi$ and electrical scalar potential $\phi(r)$ can be
written as
\begin{eqnarray}
\psi^{''}+(\frac{f'}{f}+\frac{2}{r})\psi'
+\frac{\phi^2\psi}{f^2}+\frac{2\psi}{f}=0,\label{e1}
\end{eqnarray}
and
\begin{eqnarray}
\bigg(\phi^{''}+\frac{2}{r}\phi'\bigg)\bigg(1-b\phi'^2\bigg)+b\phi'^2\phi''
-\frac{2\psi^2}{f}\phi\bigg(1-b\phi'^2\bigg)^{3/2}=0,\label{e2}
\end{eqnarray}
respectively. Here we set $m^2L^2=-2$ and a prime denotes the
derivative with respect to $r$. Obviously, the motion equation
(\ref{e2}) for the electrical scalar potential $\phi(r)$ is more
complicated than that in usual Maxwell theory. In order to solve the
nonlinear equations (\ref{e1}) and (\ref{e2}) numerically, we need
to seek the boundary condition for $\phi$ and $\psi$ near the black
hole horizon $r\sim r_H$ and at the spatial infinite
$r\rightarrow\infty$. The regularity condition at the horizon gives
the boundary conditions $\phi(r_H)=0$ and
$\psi=-\frac{3r_H}{2}\psi'$. At the spatial infinite
$r\rightarrow\infty$, the scalar filed $\psi$ and the scalar
potential $\phi$ can be approximated as
\begin{eqnarray}
\psi=\frac{\psi^{(1)}}{r}+\frac{\psi^{(2)}}{r^2}+...,\label{b1}
\end{eqnarray}
and
\begin{eqnarray}
\phi=\mu-\frac{\rho}{r}+...\;.\label{b2}
\end{eqnarray}
\begin{figure}[ht]
\begin{center}
\includegraphics[width=6.8cm]{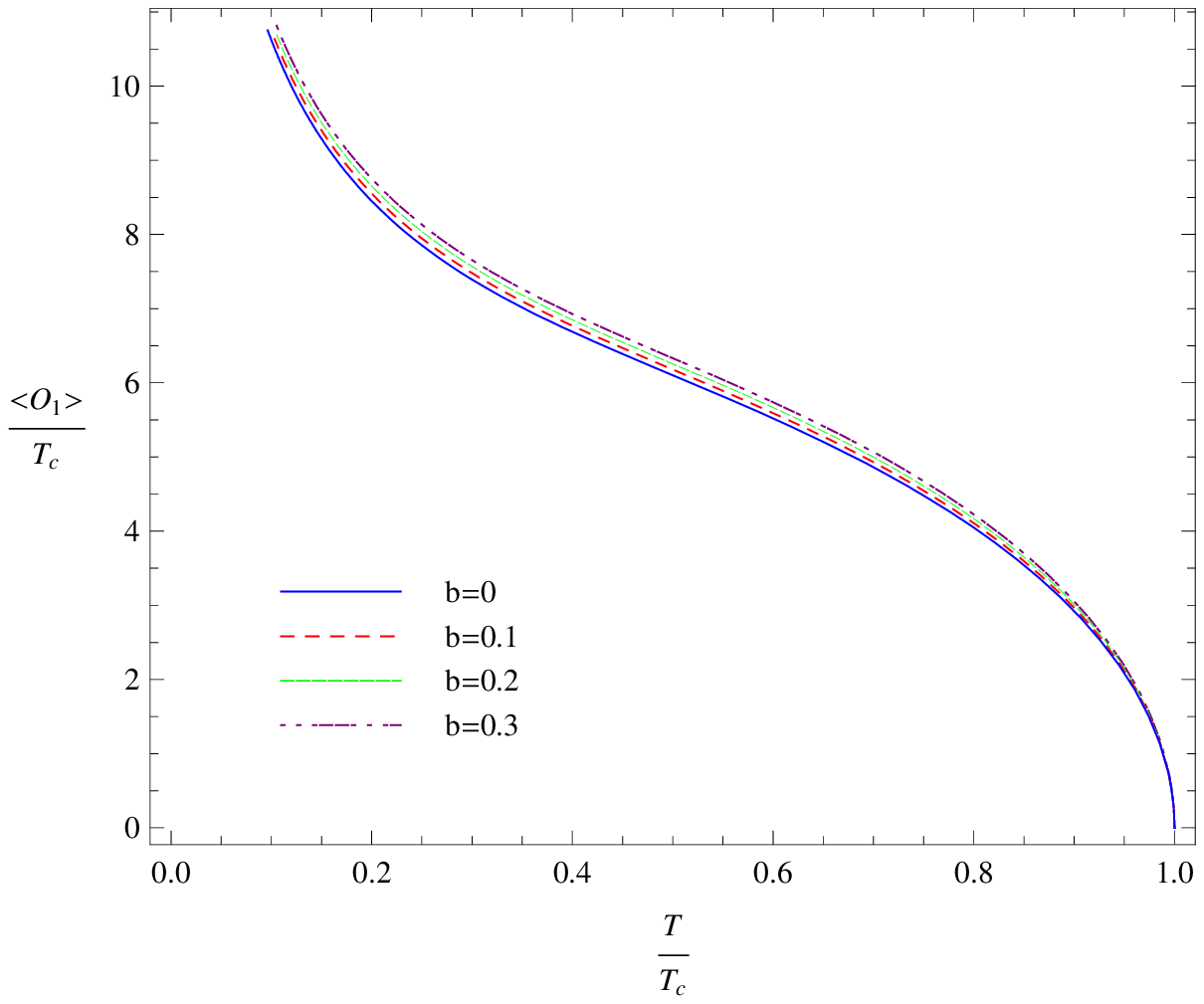}\;\;\;\;
\includegraphics[width=7cm]{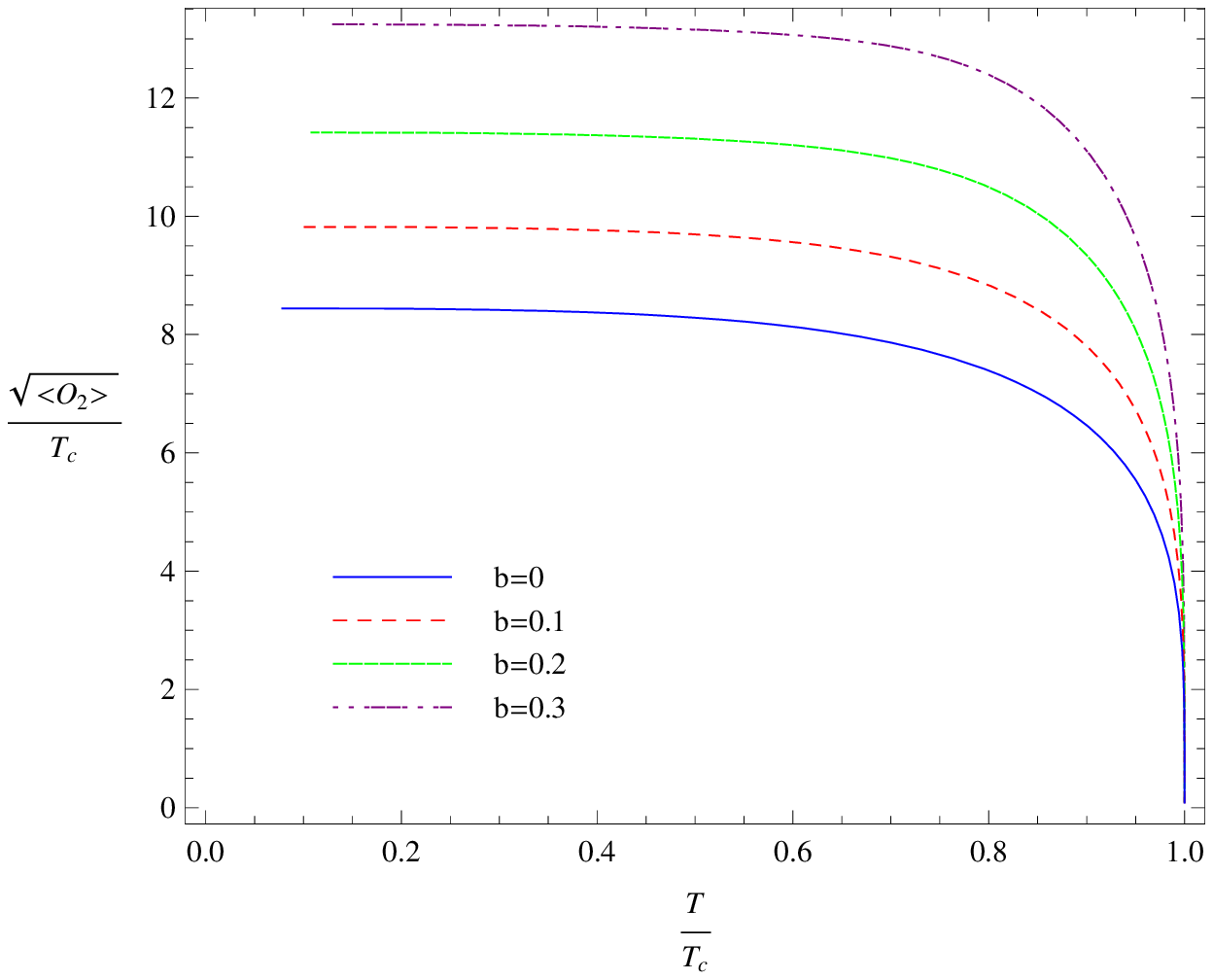}
\caption{The condensates of operators $\mathcal{O}_1$(left) and
$\mathcal{O}_2$(right) versus temperature. The condensates disappear
as $T\rightarrow T_c$. The condensate is a function of temperature
for various values of $b$. Here $b$ is the scale parameter in
Born-Infeld electrodynamics.}
\end{center}
\end{figure}
\begin{table}[ht]
\begin{tabular}[b]{|c|c|c|c|c|c|}
\hline\hline \;\;\;\;\;\; &
  \multicolumn{2}{c|}{} & \multicolumn{2}{c|}{} \\
  \;\;\;\; $b$\;\;\;\;&
  \multicolumn{2}{c|}{$\mathcal{O}_1$}& \multicolumn{2}{c|}{$\mathcal{O}_2$}\\
 \hline&&&&\\
 0&\;\;$T_c\approx0.22553
\rho^{1/2}$&$\langle\mathcal{O}_1\rangle\approx
9.31T_c(1-T/T_c)^{1/2}$&\;$T_c\approx0.11842\rho^{1/2}$
&$\langle\mathcal{O}_2\rangle\approx139.24\;T^2_c(1-T/T_c)^{1/2}$\\
&&&&\\
 \hline &&&&\\
  0.1&$T_c\approx0.22351
\rho^{1/2}$&$\langle\mathcal{O}_1\rangle\approx 9.48
  T_c(1-T/T_c)^{1/2}$&$T_c\approx0.10072\rho^{1/2}$&
   $\langle\mathcal{O}_2\rangle\approx207.36T^2_c(1-T/T_c)^{1/2}$\\
&&&&\\
 \hline&&&&\\
  0.2&$T_c\approx0.22155
\rho^{1/2}$&$\langle\mathcal{O}_1\rangle\approx 9.62
  T_c(1-T/T_c)^{1/2}$& $T_c\approx0.08566\rho^{1/2}$&$\langle\mathcal{O}_2\rangle\approx302.76T^2_c(1-T/T_c)^{1/2}$\\
&&&&\\
 \hline&&&&\\
  0.3&$T_c\approx0.21964
\rho^{1/2}$&$\langle\mathcal{O}_1\rangle\approx 9.74
  T_c(1-T/T_c)^{1/2}$& $T_c\approx0.07292\rho^{1/2}$&$\langle\mathcal{O}_2\rangle\approx432.64T^2_c(1-T/T_c)^{1/2}$\\
&&&&\\
 \hline\hline
\end{tabular}
\caption{The critical temperature and the expectation values for the
operators $\mathcal{O}_1$ and $\mathcal{O}_2$ when $T\rightarrow
T_c$ for different values of $b$.}
\end{table}
From the dual field theory, the constants $\mu$ and $\rho$ can be
interpreted as the chemical potential and charge density
respectively. The coefficients $\psi^{(1)}$ and $\psi^{(2)}$
correspond to the vacuum expectation values of the condensate
operator $\mathcal{O}$ dual to the scalar field. As in \cite{ads},
we can impose the boundary condition that either $\psi^{(1)}$ or
$\psi^{(2)}$ vanish, so that the theory is stable in the asymptotic
AdS region.

In Fig.1 we plot the variety of the condensates of operators
$\mathcal{O}_1$ and $\mathcal{O}_2$ with the Born-Infeld scale
parameter $b$. We find that at low temperatures the condensate
$\mathcal{O}_2$ has similar behaviors to the BCS theory for
different $b$. This means that there exist the holographic
superconductors when the scalar field coupled a Born-Infeld
electromagnetic field in the Schwarzschild AdS black hole. However,
we also note that the condensate $\mathcal{O}_1$ diverges as
$T\rightarrow 0$. This result is similar to that in the
Einstein-Maxwell electrodynamics \cite{Hs0} when the backreaction on
the metric was neglected. From the Fig. 1, it is easy to find that
the condensation gap increases with the Born-Infeld scale parameter
$b$. The means that the condensation gap becomes larger and the
scalar hair is formed more harder in the Born-Infeld
electrodynamics. In the table (I) we present the critical
temperature $T_c$ for the condensations and fit these condensation
curves near $T\sim T_c$ . It is easy to find that as $b$ increases
the critical temperature decreases for both condensations. The
dependence of the condensation gap and the critical temperature on
the Born-Infeld scale parameter are similar with that on the
Gauss-Bonnet term in the holographic superconductor.

\section{The electrical conductivity}

In this section we will compute the electrical conductivity by
perturbing the electric field.  Following the standard procedure in
\cite{Hs0}, we assume that the perturbation of the vector potential
has a form $\delta A_x =A_x(r)e^{-i\omega t}$.
\begin{figure}[ht]
\begin{center}
\includegraphics[width=7cm]{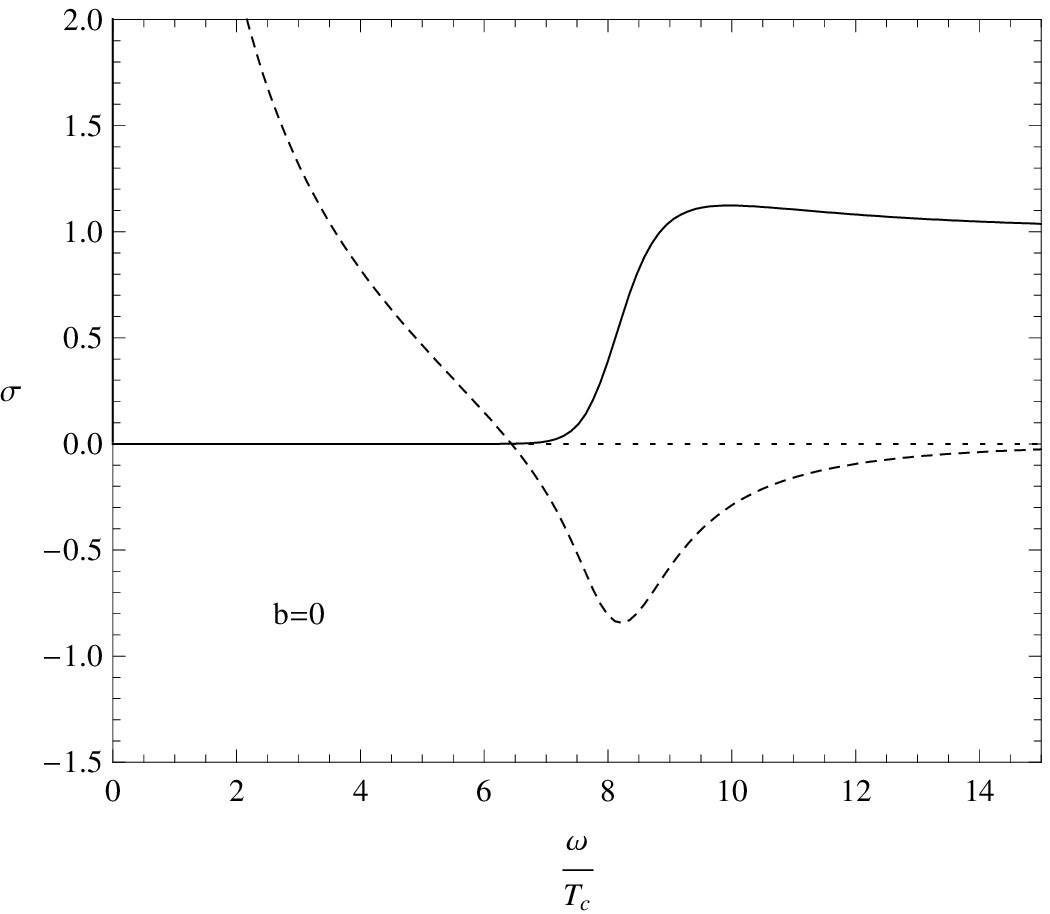}\;\;\;\;
\includegraphics[width=7cm]{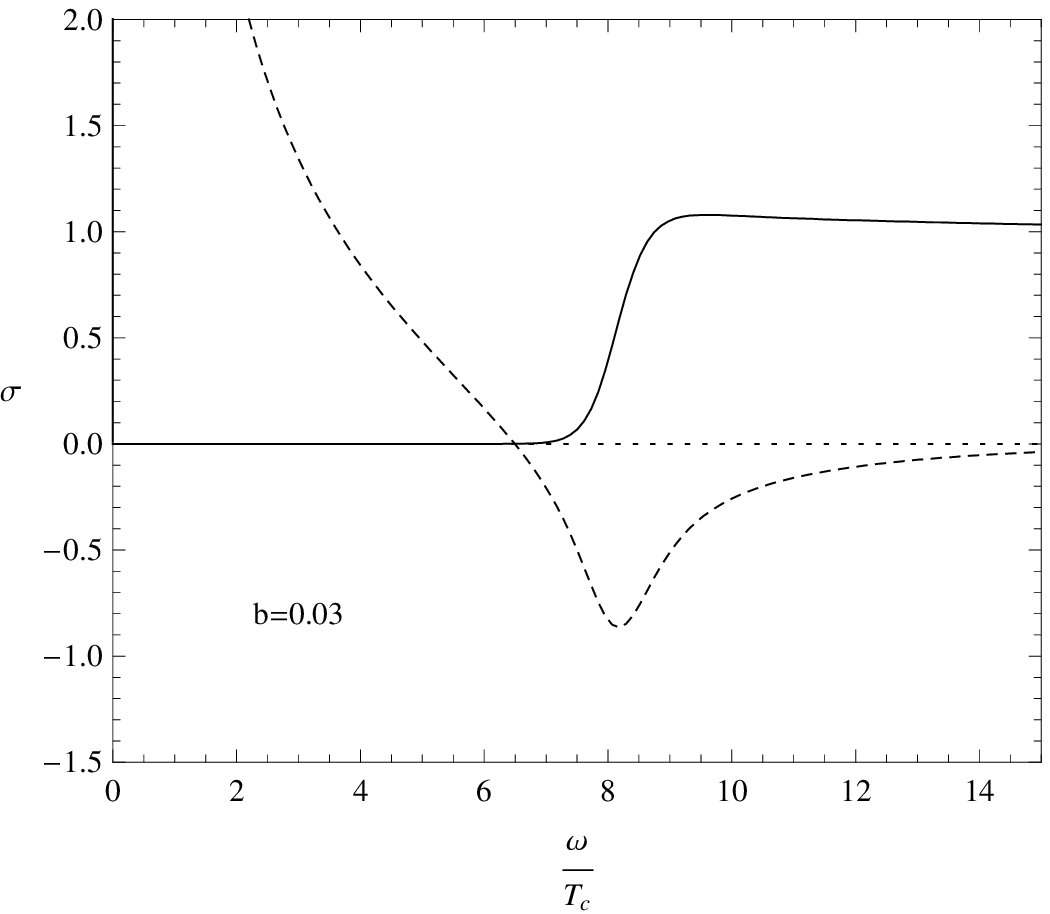}\\
\includegraphics[width=7cm]{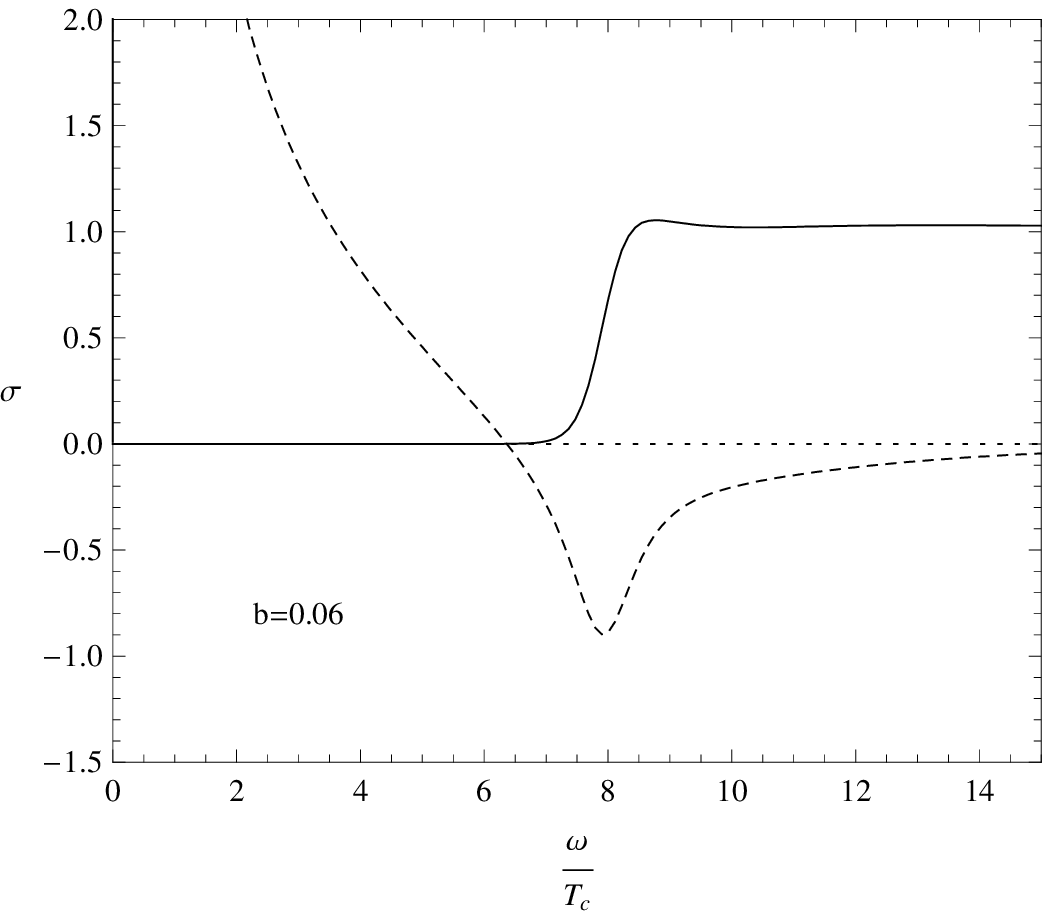}\;\;\;\;\includegraphics[width=7cm]{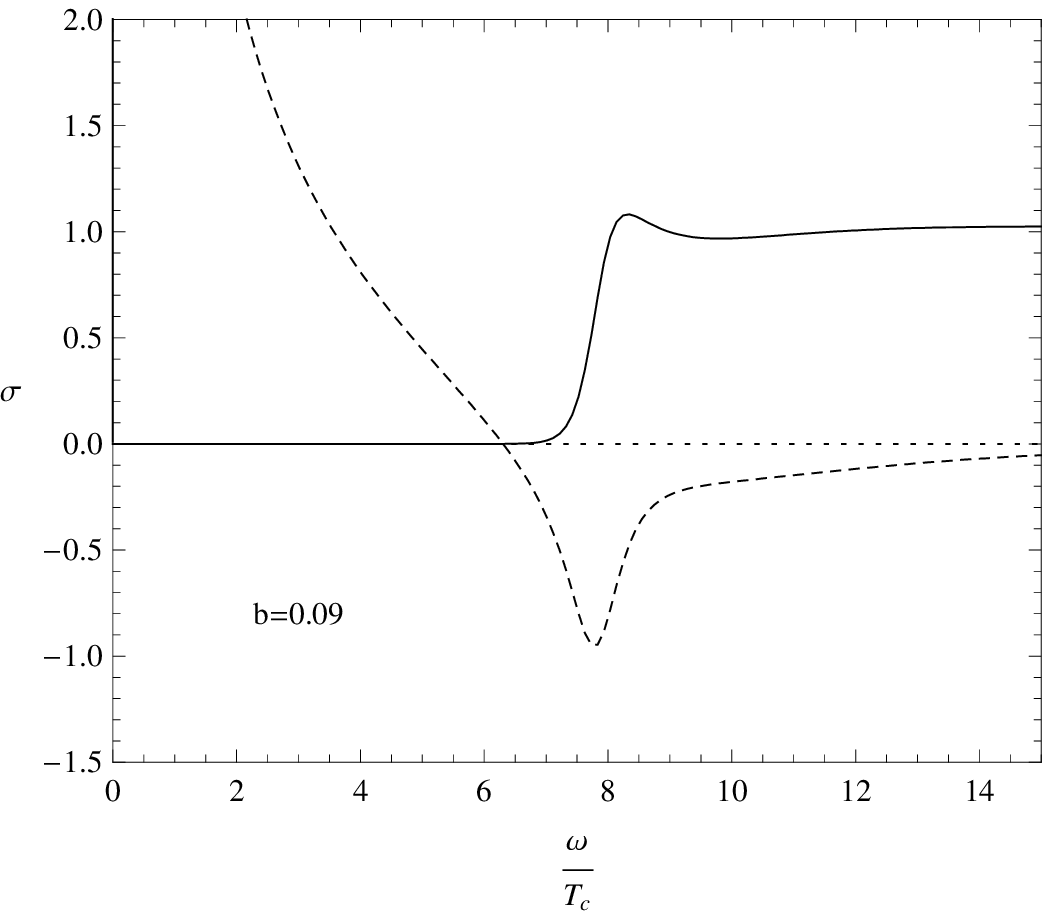}
\caption{The conductivity for operator $\langle\mathcal{O}_2\rangle$
for different values of $b$. Each plot is at low temperatures, about
$T/Tc\approx0.11$.  The solid and dashed curves denote the real part
$Re \sigma$ and the imaginary part $Im \sigma $ of conductivity,
respectively. }
\end{center}
\end{figure}
Neglecting the backreaction of the perturbational field on
background metric, one can obtain that the first-order
perturbational equation in the Born-Infeld electrodynamics has the
form
\begin{eqnarray}
\bigg[A_x^{''}+\frac{f'}{f}A_x'
+\frac{\omega^2}{f^2}A_x\bigg](1-b\phi'^2)+bA'_x\phi'\phi''-\frac{2\psi^2A_x}{f}(1-b\phi'^2)^{3/2}=0.\label{de}
\end{eqnarray}
Obviously, the above equation is more complicated than in the
Einstein-Maxwell electrodynamics because that the perturbational
field $A_x$ couples not only with the scalar field $\psi$ but also
with the scalar potential $\phi$. Fortunately, we can solve this
equation with an ingoing wave boundary condition $A_x=
f^{-\frac{i\omega }{3r_H}}$ near the black hole horizon. The general
behavior of $A_x$ at the spatial infinity can be expressed as
\begin{eqnarray}
A_x=A^{(0)}_x+\frac{A^{(1)}_x}{r}+....
\end{eqnarray}

According to the AdS/CFT, one can find \cite{Hs0} that $A^{(0)}_x$
and $A^{(1)}_x$ in the bulk corresponds to the source and the
expectation value for the current on the CFT boundary, respectively.
Applying the Ohm's law, we have the conductivity obtained in
\cite{Hs0}
\begin{eqnarray}
\sigma(\omega)=\frac{\langle J_x\rangle}{E_x}=-\frac{i\langle
J_x\rangle}{\omega A_x}=-i\frac{A^{(1)}_x}{\omega A^{(0)}_x}.
\end{eqnarray}
In Fig. 2 we plot only the frequency dependent conductivity for
operator $\langle\mathcal{O}_2\rangle$ obtained by solving the
Born-Infeld equation numerically for $b=0$, $0.03$, $0.06$ and
$0.09$. We find a gap in the conductivity with the gap frequency
$\omega_g$. As the Born-Infeld scale parameter $b$ increases the gap
frequency $\omega_g$ becomes smaller. Thus, the effects of
Born-Infeld scale parameter $b$ on the gap frequency $\omega_g$ are
different from those originated by the Gauss-Bonnet term. We also
find that there exists the similar properties of the gap frequency
for the operator $\mathcal{O}_1$.

\section{Summary}

In this paper we studied holographic superconductors in the presence
of Born-Infeld correction to Maxwell electrodynamics in
Schwarzschild AdS black hole spacetime. Considering that the
Born-Infeld theory is a main nonlinear electrodynamics, this
investigation may help to understand the effects of the nonlinear
electrodynamics on the holographic superconductors. Adopting the
probe limit, we found that the Born-Infeld scale parameter
influences the condensation formation and conductivity. The larger
Born-Infeld corrections make the scalar operator harder to condense.
Moreover, we found that the gap frequency $\omega_g$ the
conductivity becomes smaller as the parameter $b$ increases. This
effect is different from those originated by the Gauss-Bonnet
gravity. It would be of interest to generalize our study to other
nonlinear electrodynamic theories. Work in this direction will be
reported in the future.

\begin{acknowledgments}
We thank Professor Bin Wang and Dr Qiyuan Pan for their helpful
discussions and suggestions. This work was partially supported by
the National Natural Science Foundation of China under Grant
No.10675045, No.10875040 and No.10935013; 973 Program Grant No.
2010CB833004 and the Hunan Provincial Natural Science Foundation of
China under Grant No.08JJ3010. S. Chen's work was partially
supported by the National Natural Science Foundation of China under
Grant No.10875041 and the construct program of key disciplines in
Hunan Province.
\end{acknowledgments}


\vspace*{0.2cm}
\begin{thebibliography}{99}
\baselineskip=0.6 cm

\bibitem{ads1} J. M. Maldacena,  Adv. Theor. Math. Phys. {\bf2}, 231
 (1998).[hep-th/9711200].


\bibitem{ads2} E. Witten,  Adv. Theor. Math. Phys. {\bf 2}, 253 (1998).

\bibitem{ads3} S. S. Gubser, I. R. Klebanov, and A. M. Polyakov,  Phys. Lett. B {\bf 428},
105 (1998).

\bibitem{ads4} O. Aharony, S. S. Gubser, J. M. Maldacena, H. Ooguri, and Y. Oz,  Phys. Rept. {\bf323}, 183
(2000).

\bibitem{Hs0} S. A. Hartnoll, C. P. Herzog and G. T. Horowitz,  Phys. Rev. Lett. {\bf101}, 031601
(2008).

\bibitem{Hs01} S. S. Gubser,  Phys. Rev. D {\bf78}, 065034 (2008).


\bibitem{Hs02} S.A. Hartnoll, Lectures on Holographic Methods for Condensed Matter
Physics. [arXiv:0903.3246].

\bibitem{Hs03} C.P. Herzog, Lectures on Holographic
Superfluidity and Superconductivity.  J. Phys. A {\bf 42},343001
(2009)

\bibitem{Hsf0} P. Basu, A. Mukherjee and H. H. Shieh,
Phys. Rev. D {\bf 79}, 045010 (2009).

\bibitem{Hsf01}C. P. Herzog, P. K. Kovtun, and D. T. Son,  Phys. Rev. D {\bf79},
066002 (2009).

\bibitem{Hsa1} S. S. Gubser, Class. Quant. Grav. {\bf 22}, 5121
(2005).

\bibitem{bcs} J. Bardeen, L. N. Cooper and J. R. Schrieffer,  Phys. Rev. {\bf 108}, 1175 (1957).

\bibitem{a0} R. Gregory,  S. Kanno and J. Soda, Holographic Superconductors with Higher Curvature
Corrections. arXiv:0907.3203.

\bibitem{a01}H. Zeng, Z. Fan, Z. Ren, Phys. Rev. D {\bf 80}, 066001 (2009).

\bibitem{a1} S. A. Hartnoll, C. P. Herzog and G. T. Horowitz, JHEP {\bf 0812},015 (2008).

\bibitem{a2} S. S. Gubser, Phys. Rev. Lett.{\bf 101},191601 (2008).

\bibitem{a3} M. M. Roberts and S. A. Hartnoll, JHEP {\bf 08}, 035
(2008).

\bibitem{a4} S. S. Gubser and S. S. Pufu, JHEP {\bf 11}, 033 (2008).

\bibitem{a401}  Q. Y. Pan, B. Wang, E. Papantonopoulos, J. Oliveira and A.
Pavan,  arXiv:0912.2475.

\bibitem{a40}  R. Cai and H. Zhang, Holographic Superconductors with
Ho\v{r}ava-Lifshitz Black Holes, [arXiv:0911.4867].

\bibitem{a402} J. Jing, L. Wang, S. Chen, Holographic Superconductors in Ho\v{r}ava-Lifshitz
Gravity,  [arXiv:1001.2946]


\bibitem{a41} S. Sin, S. Xu and Y. Zhou, Holographic Superconductor for
a Lifshitz Fixed Point. [arXiv:0909.4857]

\bibitem{a42} E. J. Brynjolfsson, U.H. Danielsson, L. Thorlacius and T. Zingg,
Holographic Superconductors with Lifshitz Scaling. [arXiv:0908.2611]


\bibitem{a5} F. Denef and S. A. Hartnoll, Landscape of superconducting
membranes, arXiv:0901.1160.

\bibitem{a6} S. S. Gubser, C. P. Herzog, S.S. Pufu and T. Tesileanu,
Superconductors from Superstrings, Phys. Rev. Lett. {\bf 103},
141601 (2009)

\bibitem{a7} J. P. Gauntlett, J. Sonner and T. Wiseman, Holographic
superconductivity in M-Theory, Phys. Rev. Lett. {\bf 103},151601
(2009).[arXiv:0907.3796]; J. P. Gauntlett, J. Sonner and T. Wiseman,
Quantum Criticality and Holographic Superconductors in M-theory,
arXiv:0912.0512.


\bibitem{a8} S. S. Gubser, S. S. Pufu and F. D. Rocha, Quantum critical
superconductors in string theory and M-theory ,arXiv: 0908.0011.

\bibitem{a8d1} M. Ammon, J. Erdmenger, M. Kaminski and P. Kerner,
Phys. Lett. B{\bf 680} 516 (2009);

M. Ammon, J. Erdmenger, M. Kaminski and P. Kerner, JHEP {\bf0910}067
(2009).

\bibitem{GT}G. T. Horowitz and M. M. Roberts, Holographic superconductors with various condensates,
Phys. Rev. D {\bf78}, 126008 (2008).

\bibitem{a81} S. S. Gubser and A. Nellore, Ground states of holographic
superconductors.[arXiv:0908.1972].

\bibitem{a9} G. T. Horowitz and M. M. Roberts, Zero Temperature Limit of
Holographic Superconductors. JHEP {\bf 0911}, 015 (2009).
[arXiv:0908.3677].

\bibitem{a10} R. A. Konoplya and A. Zhidenko, Holographic conductivity of zero
temperature superconduc- tors, [arXiv: 0909.2138].

\bibitem{a100} T. Nishioka, S. Ryu and T.
Takayanagi,  Holographic Superconductor/Insulator Transition at Zero
Temperature,  arXiv: 0911.0962.

\bibitem{a11} S. Cremonesi, D. Melnikov and Y. Oz, ¡°Stability of
Asymptotically Schroedinger RN Black Hole and Superconductivity,
arXiv: 0911.3806 [hep-th].

\bibitem{a12} F. Aprile and J. Russo, Models of
Holographic superconductivity,  arXiv: 0912.0480.

\bibitem{a13} P. Basu, J. He, A. Mukherjee and H. Shieh, Hard-gapped Holographic
Superconductors, arXiv: 0911.4999.

\bibitem{a14} J. Sonner, Phys. Rev. D {\bf 80}, 084031 (2009).

\bibitem{a15} E. Nakano1 and W. Wen, Phys. Rev. D {\bf 78}, 046004
(2008).

\bibitem{a151}  S. Chen, L. Wang, C. Ding and J. Jing, Holographic superconductors in the AdS black hole spacetime with a global monopole, [arXiv:0912.2397]

\bibitem{a16s}  M. Ammon, J. Erdmenger, V. Grass, P. Kerner and A.
O'Bannon,  On Holographic p-wave Superfluids with Back-reaction,
arXiv:0912.3515

\bibitem{a17} K. Maeda, M. Natsuume and T. Okamura, Phys. Rev. D{\bf79},126004
(2009);

K. Maeda, M. Natsuume and T. Okamura, Phys. Rev. D{\bf81}, 026002
(2010).

\bibitem{BI} M. Born and L. Infeld, Proc. Roy. Soc. A{\bf 144}, 425 (1934).

\bibitem{BI0} G. W. Gibbons and D. A. Rasheed, Nucl. Phys. B {\bf 454}, 185
(1995).

\bibitem{BI1}E. S. Fradkin and A. A. Tseytlin, Phys. Lett. B {\bf 163}, 123
(1985).

E. Bergshoeff, E. Sezgin, C. N. Pope, and P. K. Townsend, Phys.
Lett. B {\bf 188}, 70 (1987).

\bibitem{BIbh}B. Hoffmann, Phys. Rev. {\bf47}, 877 (1935).

\bibitem{Bbh1} N. Bret\'{o}n, Class. Quantum Grav. {\bf19}, 601
(2002).

\bibitem{Bbh2} E. F. Eiroa,  Phys. Rev. D {\bf73},  043002 (2006).

\bibitem{Bbh3} O. Miskovic and R. Olea, Phys. Rev. D {\bf77}, 124048
(2008).

\bibitem{ads} I. R. Klebanov and E. Witten,  Nucl. Phys. B {\bf556}, 89 (1999).

\end{thebibliography}
\end{document}